\IfFileExists{sn-jnl.cls}
  {\documentclass[pdflatex,sn-mathphys-num]{sn-jnl}}
  {\documentclass[11pt]{article}\usepackage[margin=25mm]{geometry}}

\usepackage{graphicx}
\usepackage{amsmath,amssymb}
\usepackage{amsthm}
\usepackage{booktabs}
\usepackage{xcolor}
\usepackage{url}
\usepackage{array}
\usepackage{tikz}
\usetikzlibrary{fit}
\usetikzlibrary{shapes.misc}
\usetikzlibrary{positioning}
\usetikzlibrary{shapes.geometric}
\usetikzlibrary{arrows.meta, positioning, decorations.pathmorphing}

\tikzset{
    zxgreen/.style={circle, draw, fill=green!20, inner sep=0pt, minimum size=6mm, font=\small},
    zxred/.style={circle, draw, fill=red!20, inner sep=0pt, minimum size=6mm, font=\small},
    hbox/.style={rectangle, draw, fill=yellow, inner sep=0pt, minimum size=2.5mm},
    smalldot/.style={circle, draw, inner sep=0pt, minimum size=1.5mm}
}

\makeatletter
\@ifclassloaded{sn-jnl}{}{%
  \newcommand{\fnm}[1]{#1}\newcommand{\sur}[1]{#1}
  \newcommand{\orgdiv}[1]{#1}\newcommand{\orgname}[1]{#1}
  \newcommand{\city}[1]{#1}\newcommand{\country}[1]{#1}
  \newcommand{\bmhead}[1]{\medskip\par\noindent\textbf{#1}\par\nobreak}
  \providecommand{\backmatter}{\bigskip}
  \gdef\SNtitle{}\gdef\SNabs{}\gdef\SNkw{}\gdef\SNaff{}\gdef\SNau{}%
  \RenewDocumentCommand{\title}{o m}{\gdef\SNtitle{#2}}%
  \RenewDocumentCommand{\author}{s O{} m}{%
    \g@addto@macro\SNau{#3\textsuperscript{#2}%
      \IfBooleanT{#1}{\textsuperscript{,\,*}}\hspace{1.5em}}}%
  \NewDocumentCommand{\affil}{O{} m}{%
    \g@addto@macro\SNaff{\par\noindent\textsuperscript{#1}\,#2}}%
  \NewDocumentCommand{\email}{m}{}%
  \RenewDocumentCommand{\abstract}{m}{\gdef\SNabs{#1}}%
  \NewDocumentCommand{\keywords}{m}{\gdef\SNkw{#1}}%
  \RenewDocumentCommand{\maketitle}{}{%
    \begin{center}\LARGE\bfseries\SNtitle\par\end{center}
    \begin{center}\normalsize\SNau\par\end{center}
    \begingroup\footnotesize\SNaff
      \par\noindent\textsuperscript{*}\,corresponding author\par\endgroup
    \medskip
    \begin{quotation}\noindent\textbf{Abstract}\par\SNabs
      \end{quotation}
    \medskip}%
  \newtheoremstyle{thmstyleone}{}{}{\itshape}{}{\bfseries}{.}{.5em}{}%
  \newtheoremstyle{thmstyletwo}{}{}{\normalfont}{}{\itshape}{.}{.5em}{}%
}
\makeatother

\theoremstyle{thmstyleone}
\newtheorem{theorem}{Theorem}

\theoremstyle{thmstyletwo}
\newtheorem{remark}{Remark}

\ifdefined\hypersetup\hypersetup{bookmarksdepth=3}\fi
\providecommand{\texorpdfstring}[2]{#1}

\newcommand{\DOI}{21910179}

\newif\ifdecide\decidetrue

\begin{document}

\title[ZX-Calculus Optimization of Solovay--Kitaev Circuits]{Numerical
Evaluation of ZX-Calculus Optimization for Solovay--Kitaev Quantum Circuit
Synthesis}

\author[1]{\fnm{Dulari} \sur{De Silva}}

\author[1,2,3]{\fnm{Anuradha} \sur{Mahasinghe}}

\author[3,4]{\fnm{Chon-Fai} \sur{Kam}}\email{dubussygauss@gmail.com}

\author[2,3]{\fnm{Kaushika} \sur{De Silva}}

\author[3,5]{\fnm{Frederic} \sur{Cadet}}

\author*[2]{\fnm{Jingbo} \sur{Wang}}\email{jingbo.wang@uwa.edu.au}

\affil[1]{\orgdiv{Department of Mathematics}, \orgname{University of Colombo},
\city{Colombo}, \country{Sri Lanka}}

\affil[2]{\orgdiv{School of Physics, Mathematics and Computing},
\orgname{The University of Western Australia}, \city{Perth},
\country{Australia}}

\affil[3]{\orgname{University Paris City and University of Reunion},
\city{Paris}, \country{France}}

\affil[4]{\orgdiv{Dipartimento di Fisica e Chimica Emilio Segr\`e},
\orgname{Universit\`a degli Studi di Palermo}, \city{Palermo},
\country{Italy}}

\affil[5]{\orgname{PEACCEL, AI for Biologics}, \city{Paris},
\country{France}}

\abstract{\unboldmath Fault-tolerant architectures implement non-Clifford $T$ gates
through magic-state distillation, so the $T$-count of a synthesized circuit
dominates its physical cost. The Solovay--Kitaev algorithm approximates any
single-qubit unitary from a finite gate set with a sequence length that grows
only polylogarithmically in the inverse target error, but it optimizes for
numerical convergence rather than circuit economy, and its output carries
structural redundancy that a gate-level compiler cannot see. We report a
measurement of what diagrammatic post-processing recovers from that
redundancy. Twelve hundred random single-qubit targets, spanning the three
Pauli rotation families and the general gate $U(\theta,\phi,\lambda)$, are
synthesized over Clifford$+T$ at three recursion depths, translated into
graph-like ZX-diagrams, simplified by automated rewriting, and extracted back
to circuits. Post-processing removes 26.6--30.1\% of the total gate count and
18.5--22.2\% of the $T$-count. The absolute saving grows with recursion depth,
from about 60 to about 1600 gates, while the fractional saving does not: it
rises slightly from the shallowest setting and is then flat across a
twenty-five-fold change in circuit length, and by the deepest setting the four
target families are no longer distinguishable from one another. Because the
rewrite rules preserve the implemented linear map,
the approximation error is unchanged. The compile-time cost of the rewriting
layer, by contrast, grows sharply with depth and comes to dominate the
synthesis itself.}


\maketitle

\section{Introduction}\label{sec:intro}

Quantum computing offers the potential to solve certain computational problems
more efficiently than classical computers by exploiting superposition and
entanglement, and landmark algorithms for integer factorization, database
search and quantum simulation have established the theoretical
advantage~\cite{nielsen2010quantum}. Realizing that advantage at scale remains
difficult, because quantum systems are susceptible to noise, decoherence and
operational error~\cite{kitaev1997quantum}. Fault-tolerant quantum computation
addresses this by encoding logical qubits into error-correcting codes and
implementing logical operations so that physical errors are
suppressed~\cite{kitaev2002classical}.

Within most fault-tolerant frameworks the cost of a logical operation is
sharply non-uniform. Clifford operations, including the Hadamard, Pauli and
controlled-NOT gates, can be implemented with comparatively low
overhead~\cite{aaronson2004improved}. Universality requires a non-Clifford
operation, most commonly the $T$ gate, and a fault-tolerant $T$ gate is
typically obtained by magic-state distillation, which consumes many physical
qubits and many rounds of error correction. The $T$-count of a circuit is
therefore the primary engineering cost metric for fault-tolerant
compilation~\cite{amy2014polynomial,kissinger2019reducing}.

Because hardware supports only a finite set of elementary operations, an
arbitrary unitary must be approximated by a sequence drawn from a universal
gate set. The Solovay--Kitaev theorem guarantees that any single-qubit unitary
can be approximated to arbitrary precision from a finite universal set, with
an approximating sequence whose length grows only polylogarithmically in the
inverse of the target error~\cite{kitaev1997quantum,nielsen2001quantum,
dawson2005solovay,mahasinghe2020solovay}. The construction is recursive: each
level corrects the residual error of the level below by factoring it into a
group commutator. That recursion is what makes the theorem general, and it is
also what makes its output redundant, since the nested commutators reintroduce
gates that cancel against one another once the sequence is read as a whole.

Diagrammatic optimization offers a way to see those cancellations. The
ZX-calculus represents a quantum computation not as an ordered list of gates
but as a graph of spiders and edges, together with a set of sound rewrite
rules~\cite{coecke2011interacting,duncan2020graph}. Rewriting is
semantics-preserving, so any simplification it performs leaves the implemented
linear map unchanged, and because the representation discards the temporal
ordering of the circuit it can merge phase nodes that a linear, gate-by-gate
compiler cannot bring together. Automated implementations such as
PyZX~\cite{kissinger2019pyzx} apply these rules to a normal form and extract a
circuit at the end.

Synthesis and optimization are usually studied as separate stages, and the
question of how much a diagrammatic optimizer recovers from a
Solovay--Kitaev circuit specifically has received little empirical attention.
This paper reports a measurement of that composition. Twelve hundred random
single-qubit targets, spanning the three Pauli rotation families and the general
gate $U(\theta,\phi,\lambda)$, are synthesized at three recursion depths,
post-processed with automated ZX rewriting, and compared on total gate count,
$T$-count and compile time.

The contribution is a measurement, not a method. Its interest lies in three
features of the result. First, the fraction of the circuit that rewriting
removes is close to constant in the recursion depth: it rises slightly from the
shallowest setting and is then flat, even though the circuit grows
twenty-five-fold and the approximation error falls by nearly two orders of
magnitude across the range tested. The absolute saving grows; the proportional
saving does not. Second, that fraction is also close to constant across target
families, and increasingly so with depth: at the deepest setting the four
families agree to within a quarter of a percentage point, and the spread across
individual targets falls by a factor of nearly three. Third, the compile-time
cost of obtaining the saving grows steeply, so the composition has an operating
range rather than a uniform benefit.

Section~\ref{sec:related} reviews synthesis and diagrammatic optimization and
positions this work against near-optimal synthesis.
Section~\ref{sec:background} gives the necessary background.
Section~\ref{sec:method} describes the pipeline.
Section~\ref{sec:results} reports and analyses the measurements.
Section~\ref{sec:conclusion} concludes.

\section{Related work}\label{sec:related}

\subsection{Synthesis over \texorpdfstring{Clifford$+T$}{Clifford+T}}\label{sec:related-synth}

In fault-tolerant paradigms such as those built on surface codes, operations
are restricted to a discrete set of protected
primitives~\cite{nielsen2010quantum,nielsen2001quantum}. The most widely
adopted is the Clifford$+T$ library: the Clifford generators together with the
non-Clifford $T$ gate, a $\pi/4$ phase rotation. Clifford operations are
comparatively cheap~\cite{aaronson2004improved} while $T$ gates require
distillation~\cite{kitaev2002classical}, which is why $T$-count is the metric
of record~\cite{amy2014polynomial,kissinger2019reducing}.

The synthesis of arbitrary elements of $SU(2)$ was first settled by the
Solovay--Kitaev theorem~\cite{kitaev1997quantum,nielsen2001quantum,
ozols2009solovay,roje2019solovay}, which produces a sequence of length
$O(\log^{c}(1/\epsilon))$ for a constant $c$ determined by the generating
set~\cite{dawson2005solovay,mahasinghe2020solovay}.

It is important to state plainly where that leaves Solovay--Kitaev relative to
current practice, because the comparison governs how the present measurement
should be read. Solovay--Kitaev is not the method a production compiler uses
for a $z$-rotation. Number-theoretic synthesis is far more economical:
Kliuchnikov, Maslov and Mosca gave an asymptotically optimal
procedure~\cite{kmm2013}, and Ross and Selinger gave an ancilla-free algorithm
that is optimal given a factoring oracle and near-optimal without one,
achieving in the typical case a $T$-count of $3\log_{2}(1/\epsilon) +
O(\log\log(1/\epsilon))$~\cite{ross-selinger2016}. At the accuracies
considered in this paper that leading term is a few tens of $T$ gates, one to
two orders of magnitude below the sequences Solovay--Kitaev produces.

This paper therefore does not propose Solovay--Kitaev as a competitive
synthesis route, and the reductions reported below should not be read as
competitive $T$-counts. Solovay--Kitaev remains of interest for a different
reason: it is the general construction, applying to arbitrary elements of
$SU(2)$ and to arbitrary inverse-closed generating sets, where the
number-theoretic methods are specialized to $z$-rotations over Clifford$+T$.
The question this paper measures is how much recoverable structural redundancy
a general recursive group-commutator construction leaves behind, which is a
question about the output of that construction rather than a proposal to use
it.

\subsection{Diagrammatic circuit optimization}\label{sec:related-zx}

The ZX-calculus, introduced by Coecke and
Duncan~\cite{coecke2011interacting}, is a graphical language for quantum
processes. It replaces an ordered gate list with a spatial arrangement of
nodes called spiders, labelled by phases and belonging to the $Z$ or $X$
basis~\cite{duncan2020graph,coecke2018picturing}. The calculus is complete for
stabilizer quantum mechanics~\cite{backens2014zx}, and its rewrite rules
preserve the implemented linear map up to a global scalar. Automated
frameworks, principally PyZX~\cite{kissinger2019pyzx}, drive these rules to a
graph-like normal form and then extract a
circuit~\cite{duncan2020graph,kissinger2019cnot}. Non-diagrammatic approaches
to the same objective include $T$-depth optimization by matroid
partitioning~\cite{amy2014polynomial}, automated optimization of circuits with
continuous parameters~\cite{nam2018automated}, and treating $T$ gates as
$\pi/4$ Pauli rotations~\cite{zhang2019optimizing}. This paper compares
against unoptimized Solovay--Kitaev output only; a comparison against those
optimizers is left to future work, and is noted as a limitation in
Section~\ref{sec:limitations}.

\section{Background}\label{sec:background}

\subsection{The Solovay--Kitaev construction}\label{sec:bg-sk}

\begin{theorem}[Solovay--Kitaev]\label{thm:sk}
Let $G \subset SU(2)$ be a finite set, closed under inverses
($g \in G \Rightarrow g^{-1} \in G$), which generates a subgroup dense in
$SU(2)$. For a target $U \in SU(2)$ and a precision $\epsilon > 0$ there
exists a finite sequence $S = g_1 g_2 \cdots g_l$ of elements of $G$ with
\begin{equation}
d(U, S) < \epsilon ,
\end{equation}
and the required length obeys
\begin{equation}\label{eq:polylog}
l = O\!\left(\log^{c}\!\left(\tfrac{1}{\epsilon}\right)\right).
\end{equation}
\end{theorem}

\begin{remark}
The hypothesis is that the subgroup \emph{generated by} $G$ is dense; a finite
set cannot itself be dense in $SU(2)$.
\end{remark}

The generating set used here is
\begin{equation}\label{eq:gateset}
G = \{H,\, T,\, T^{\dagger},\, S,\, S^{\dagger},\, Z\},
\end{equation}
which is closed under inverses, since $H$ and $Z$ are self-inverse and the
remaining elements appear with their adjoints. These gates are chosen because
they map onto fault-tolerant hardware built around magic-state
distillation~\cite{nielsen2010quantum}.

The construction is recursive. Fix an initial threshold $\epsilon_0$ and
require, for stability of the recursion, that
\begin{equation}\label{eq:stability}
\epsilon_k^{2} < \epsilon_{k+1}.
\end{equation}
At the base of the recursion a precomputed net of all products of elements of
$G$ up to a fixed length $l_0$ is searched for a root approximation
$U_0 \in G^{l_0}$ minimizing the distance to
$U$~\cite{dawson2005solovay,ozols2009solovay}:
\begin{equation}\label{eq:base}
\lVert U - U_0 \rVert < \epsilon_0^{2}.
\end{equation}

The residual is tracked by the difference operator
$\Delta_1 = U U_0^{\dagger}$. Since $U_0$ is unitary,
\begin{equation}\label{eq:delta1}
\lVert \Delta_1 - I \rVert
= \lVert (U - U_0) U_0^{\dagger} \rVert
= \lVert U - U_0 \rVert < \epsilon_0^{2} < \epsilon_1 ,
\end{equation}
so $\Delta_1$ lies in a ball about the identity and may be treated as a new
target. Being close to the identity, it represents an infinitesimal rotation
of the Bloch sphere and can be written as a group commutator,
\begin{equation}\label{eq:comm}
\Delta_1 = A B A^{\dagger} B^{\dagger}.
\end{equation}
The factors are obtained through the Lie algebra: any element of $SU(2)$
corresponds to a rotation vector $\vec r$ under
\begin{equation}\label{eq:expmap}
u(\vec r) = \exp\!\left(-i\,\frac{\vec r \cdot \vec\sigma}{2}\right),
\end{equation}
with $\vec\sigma = (\sigma_x, \sigma_y, \sigma_z)$ the Pauli vector. Because
$\lVert \Delta_1 - I \rVert < \epsilon_0^2$, the corresponding vector
satisfies $|\vec r| \approx \lVert \Delta_1 - I\rVert$, and the algorithm
solves for $\vec a, \vec b$ with $\vec a \times \vec b \approx \vec r$ and
$|\vec a|, |\vec b| < \epsilon_0$~\cite{dawson2005solovay}. Mapping these back
gives $A, B$ with $\lVert A - I\rVert, \lVert B - I \rVert < \epsilon_0$, and
the base net supplies discrete approximations $A_1, B_1$ with
\begin{equation}\label{eq:AB}
\lVert A - A_1 \rVert < \epsilon_0^{2}, \qquad
\lVert B - B_1 \rVert < \epsilon_0^{2}.
\end{equation}
The correction is their commutator,
\begin{equation}\label{eq:C}
U_1 = A_1 B_1 A_1^{\dagger} B_1^{\dagger},
\end{equation}
a word of length at most $4l_0$, and
$\lVert U U_0^{\dagger} - U_1 \rVert < \epsilon_1$. The approximation after
one level of correction is the product $U_1 U_0$, of length at most $5l_0$.

\begin{remark}
Equation~\eqref{eq:C} defines $U_1$ as the correction alone, so that the
compiled approximant below is a product in which $U_0$ appears exactly once.
\end{remark}

At depth $k$ the residual is tracked by
\begin{equation}\label{eq:deltak}
\Delta_k = \Delta_{k-1} U_{k-1}^{\dagger}
= U U_0^{\dagger} U_1^{\dagger} \cdots U_{k-1}^{\dagger},
\end{equation}
whose distance to the identity is the distance between the target and the
compiled word,
\begin{equation}\label{eq:deltak2}
\lVert \Delta_k - I \rVert
= \lVert U - U_{k-1} U_{k-2} \cdots U_1 U_0 \rVert
< \epsilon_{k-1}^{2} < \epsilon_k .
\end{equation}
The compiled circuit is the ordered product
\begin{equation}\label{eq:approx}
U_{\mathrm{approx}} = U_k U_{k-1} \cdots U_1 U_0 ,
\qquad
\lVert U - U_{\mathrm{approx}} \rVert < \epsilon_k^{2} .
\end{equation}
Each correction $U_j$ with $j \ge 1$ is a commutator of four level-$(j-1)$
approximations and so has length $4 \cdot 5^{\,j-1} l_0$, while $U_0$ has
length $l_0$. The total length of the compiled circuit is therefore
\begin{equation}\label{eq:length}
L = l_0 + 4 l_0 \sum_{m=0}^{k-1} 5^{m}
  = l_0 + 4 l_0 \cdot \frac{5^{k} - 1}{4}
  = 5^{k} l_0 .
\end{equation}
By the Shrinking Lemma the error contracts as
\begin{equation}\label{eq:shrink}
\epsilon_k^{2} = \frac{(s\epsilon_0)^{(3/2)^{k}}}{s^{2}} = \epsilon ,
\end{equation}
where $s$ is the Shrinking-Lemma constant of the generating set, the factor by
which one level of commutator factorization contracts the residual
norm~\cite{dawson2005solovay}. Its value depends on $G$ and enters
the bound below only through additive constants inside the logarithms, so the
asymptotic statement does not require it to be evaluated. Taking logarithms,
\begin{equation}\label{eq:kfromeps}
\left(\frac{3}{2}\right)^{k}
= \frac{\log(1/s^{2}\epsilon)}{\log(1/s\epsilon_0)} ,
\end{equation}
and the balance between the fivefold growth of the word and the
$3/2$-exponential contraction of the error fixes the exponent
in~\eqref{eq:polylog},
\begin{equation}\label{eq:c}
c = \frac{\log 5}{\log(3/2)} \approx 3.97 .
\end{equation}
Substituting~\eqref{eq:kfromeps} and~\eqref{eq:c} into~\eqref{eq:length},
\begin{equation}\label{eq:Lfinal}
L = 5^{k} l_0
= \left(\frac{\log(1/s^{2}\epsilon)}{\log(1/s\epsilon_0)}\right)^{c} l_0
= O\!\left(\log^{c}(1/\epsilon)\right),
\end{equation}
so the circuit size is polylogarithmic in the precision. Because the
commutators are nested, however, the raw output carries a large amount of
structural redundancy, and recovering it is the subject of this paper.

\subsection{ZX-diagrams and graph-like form}\label{sec:bg-zx}

A ZX-diagram is built from wires and structural generators called spiders,
which map directly onto linear operations and tensors. Wires on the left are
inputs and wires on the right are outputs, and diagrams are compositional: the
outputs of one may be joined to the inputs of another. Spiders represent states
and unitaries in both the computational basis $\{|0\rangle, |1\rangle\}$ and
the Hadamard-transformed basis $\{|+\rangle, |-\rangle\}$~\cite{backens2014zx}.
There are two kinds:

\begin{enumerate}
\item \emph{$Z$-spiders}, drawn as green circles, act in the computational
basis and carry a phase~\cite{duncan2020graph}.
\item \emph{$X$-spiders}, drawn as red circles, act in the Hadamard-transformed
basis~\cite{duncan2020graph}.
\end{enumerate}

\begin{figure}[!ht]
\centering
  \begin{minipage}[b]{0.48\textwidth}
    \centering
    \begin{tikzpicture}[scale=0.6, baseline=(g.center)]
        \node[circle, draw, fill=green!40, minimum size=6mm] (g) at (0,0) {$\alpha$};
        \draw (-1.5,0.6) .. controls (-0.8,0.3) .. (g);
        \draw (-1.5,-0.6) .. controls (-0.8,-0.3) .. (g);
        \draw (g) .. controls (0.8,0.3) .. (1.5,0.6);
        \draw (g) .. controls (0.8,-0.3) .. (1.5,-0.6);
        \foreach \y in {-0.1, 0, 0.1} {
            \fill (-1.5,\y) circle (0.5pt);
            \fill (1.5,\y) circle (0.5pt);
        }
    \end{tikzpicture}
    \quad
    \scalebox{0.75}{$|0\dots0\rangle \langle 0\dots0| + e^{i\alpha}|1\dots1\rangle \langle 1\dots1|$}
    \caption{Z-spider linear map.}
    \label{fig:z-spider-map}
  \end{minipage}
  \hfill
  \begin{minipage}[b]{0.48\textwidth}
    \centering
    \begin{tikzpicture}[scale=0.6, baseline=(r.center)]
        \node[circle, draw, fill=red!40, minimum size=6mm] (r) at (0,0) {$\alpha$};
        \draw (-1.5,0.6) .. controls (-0.8,0.3) .. (r);
        \draw (-1.5,-0.6) .. controls (-0.8,-0.3) .. (r);
        \draw (r) .. controls (0.8,0.3) .. (1.5,0.6);
        \draw (r) .. controls (0.8,-0.3) .. (1.5,-0.6);
        \foreach \y in {-0.1, 0, 0.1} {
            \fill (-1.5,\y) circle (0.5pt);
            \fill (1.5,\y) circle (0.5pt);
        }
    \end{tikzpicture}
    \quad
    \scalebox{0.75}{$|+\dots+\rangle \langle +\dots+| + e^{i\alpha}|-\dots-\rangle \langle -\dots-|$}
    \caption{X-spider linear map.}
    \label{fig:x-spider-map}
  \end{minipage}
\end{figure}

In Figure~\ref{fig:z-spider-map} the binary values inside the bra determine the
number of input wires and those inside the ket the number of output wires; the
same reading applies to Figure~\ref{fig:x-spider-map} on the $|\pm\rangle$ basis.
Simple states and rotations follow directly.

\begin{figure}[!ht]
\centering
\[
\begin{array}{cc}
    \begin{array}{c@{\quad}c}
    \begin{tikzpicture}[scale=0.7, baseline=(g.center)]
    \node[circle, draw, fill=green!40, minimum size=5mm] (g) at (0,0) {$\alpha$};
    \draw (g) -- (1.2,0);
    \end{tikzpicture}
    &
    |0\rangle + e^{i\alpha}|1\rangle
    \end{array}
    & 
    \quad \quad
    \begin{array}{c@{\quad}c}
    \begin{tikzpicture}[scale=0.7, baseline=(r.center)]
    \node[circle, draw, fill=red!40, minimum size=5mm] (r) at (0,0) {$\alpha$};
    \draw (r) -- (1.2,0);
    \end{tikzpicture}
    &
    |+\rangle + e^{i\alpha}|-\rangle
    \end{array} \\
    \text{\small (a) Single-qubit state (Z-basis)} & \quad \quad \text{\small (b) Single-qubit state (X-basis)} \\
    [2em] 
    
    \begin{array}{c@{\quad}c}
    \begin{tikzpicture}[scale=0.7, baseline=(g.center)]
    \node[circle, draw, fill=green!40, minimum size=5mm] (g) at (0,0) {$\alpha$};
    \draw (-1.2,0) -- (g) -- (1.2,0);
    \end{tikzpicture}
    &
    | 0 \rangle \langle 0 | + e^{i\alpha} | 1 \rangle \langle 1 | = Z_{\alpha}
    \end{array}
    & 
    \quad \quad
    \begin{array}{c@{\quad}c}
    \begin{tikzpicture}[scale=0.7, baseline=(r.center)]
    \node[circle, draw, fill=red!40, minimum size=5mm] (r) at (0,0) {$\alpha$};
    \draw (-1.2,0) -- (r) -- (1.2,0);
    \end{tikzpicture}
    &
    |+ \rangle \langle + | + e^{i\alpha} |- \rangle \langle - | = X_{\alpha}
    \end{array} \\
    \text{\small (c) $Z_{\alpha}$ rotation} & \quad \quad \text{\small (d) $X_{\alpha}$ rotation} \\
\end{array}
\]
\caption{Diagrammatic representations of foundational ZX-calculus spider components: (a) Single-qubit state preparation in the green $Z$-basis; (b) Single-qubit state preparation in the red $X$-basis; (c) Green $Z_{\alpha}$ rotation operator with single input and single output; (d) Red $X_{\alpha}$ rotation operator with single input and single output.}
\label{fig:zx_spiders}
\end{figure}
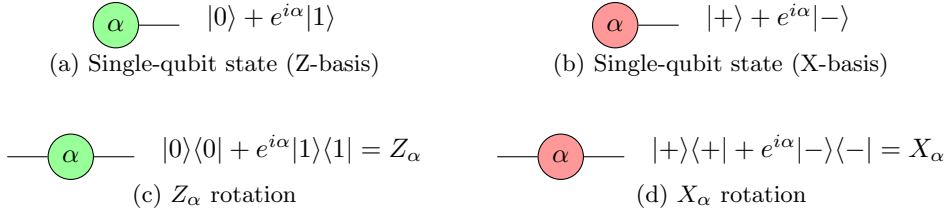

The gates that matter for Clifford$+T$ synthesis all have compact diagrammatic
forms, including Pauli-$X$, Pauli-$Z$, CNOT, the phase gate $S$, the $T$ gate,
CZ and the Hadamard~\cite{duncan2020graph,backens2014zx}. In particular the two
non-trivial generators used in this paper are single green spiders: $S$ at
phase $\pi/2$ and $T$ at phase $\pi/4$.

\begin{figure}[!ht]
\centering
\[
\begin{array}{cc}
    \begin{array}{c@{\quad}c}
    \begin{tikzpicture}[scale=0.7, baseline=(r.center)]
      \node[draw, circle, fill=red!20, minimum size=6mm, inner sep=1pt] (r) at (0,0) {$\pi$};
      \draw[-] (-1.2,0) -- (r);
      \draw[-] (r) -- (1.2,0);
    \end{tikzpicture}
    &
    |+\rangle \langle+| + e^{i\pi}|-\rangle \langle-|
    \end{array}
    & 
    \quad \quad
    \begin{array}{c@{\quad}c}
    \begin{tikzpicture}[scale=0.7, baseline=(g.center)]
      \node[draw, circle, fill=green!20, minimum size=6mm, inner sep=1pt] (g) at (0,0) {$\pi$};
      \draw[-] (-1.2,0) -- (g);
      \draw[-] (g) -- (1.2,0);
    \end{tikzpicture}
    &
    |0\rangle \langle0| + e^{i\pi}|1\rangle \langle 1 |
    \end{array} \\
    \text{\small (a) Pauli-X gate} & \quad \quad \text{\small (b) Pauli-Z gate} \\
    [2.5em] 
    
    \begin{array}{c@{\quad}c}
    \begin{tikzpicture}[baseline=0.35cm, scale=0.7] 
        \draw (0,1) -- (2,1); 
        \draw (0,0) -- (2,0);
        \draw (1,1) -- (1,0); 
        \node[draw, circle, fill=green!20, inner sep=2pt, minimum size=3mm] at (1,1) {}; 
        \node[draw, circle, fill=red!20, inner sep=2pt, minimum size=3mm] at (1,0) {}; 
    \end{tikzpicture}
    &
    \text{CNOT}
    \end{array}
    & 
    \quad \quad
    \begin{array}{c@{\quad}c}
    \begin{tikzpicture}[baseline=(g.center), scale=0.7]
        \draw (0,0) -- (2,0);           
        \node[draw, circle, fill=green!20, inner sep=1pt, minimum size=6mm] (g) at (1,0) {$\pi/2$}; 
    \end{tikzpicture}
    &
    |0\rangle \langle0| + e^{i\pi/2}|1\rangle \langle 1 |
    \end{array} \\
    \text{\small (c) CNOT gate} & \quad \quad \text{\small (d) S gate} \\
    [2.5em]
    
    \begin{array}{c@{\quad}c}
    \begin{tikzpicture}[baseline=(g.center), scale=0.7]
        \draw (0,0) -- (2,0);           
        \node[draw, circle, fill=green!20, inner sep=1pt, minimum size=6mm] (g) at (1,0) {$\pi/4$}; 
    \end{tikzpicture}
    &
    |0\rangle \langle0| + e^{i\pi/4}|1\rangle \langle 1 |
    \end{array}
    & 
    \quad \quad
    \begin{array}{c@{\quad}c}
    \begin{tikzpicture}[baseline=0.35cm, scale=0.7] 
        \draw (0,1) -- (2,1);
        \draw (0,0) -- (2,0); 
        \draw [blue, thick](1,1) -- (1,0); 
        \node[draw, circle, fill=green!20, inner sep=2pt, minimum size=3mm] at (1,1) {}; 
        \node[draw, circle, fill=green!20, inner sep=2pt, minimum size=3mm] at (1,0) {}; 
    \end{tikzpicture}
    &
    \text{CZ}
    \end{array} \\
    \text{\small (e) T gate} & \quad \quad \text{\small (f) CZ gate} \\
    [2.5em]

    \multicolumn{2}{c}{
        \begin{array}{c@{\qquad}c}
        \begin{tikzpicture}[baseline=-0.5ex, scale=0.7]
          \draw (-1.5,0) -- (1.5,0);
          \draw[fill=yellow!60, draw=black] (-0.3,-0.3) rectangle (0.3,0.3);
          \node at (0,0) {\small $H$};
        \end{tikzpicture}
        &
        |+ \rangle \langle 0| + |-\rangle \langle 1|
        \end{array}
    } \\
    \multicolumn{2}{c}{\text{\small (g) Hadamard gate}} \\
\end{array}
\]
\caption{ZX-calculus representations and matrix mappings for basic quantum gates: (a) Pauli-X ($\pi$-phase red spider); (b) Pauli-Z ($\pi$-phase green spider); (c) CNOT (green control linked to a red target); (d) S gate ($\pi/2$ green spider); (e) T gate ($\pi/4$ green spider); (f) CZ gate (two green spiders with a blue edge); (g) Hadamard gate (yellow box).}
\label{fig:zx_quantum_gates}
\end{figure}
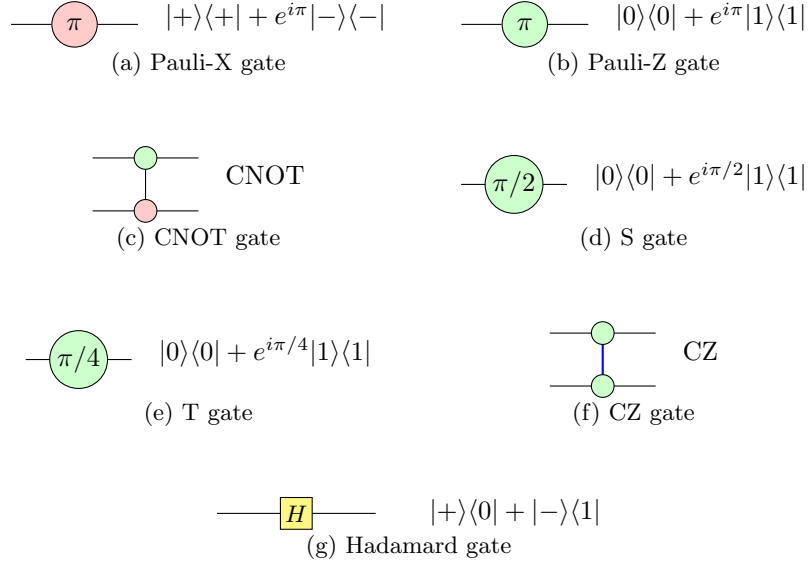

Unlike ordinary circuit notation, the ZX-calculus is not only a drawing
convention: it comes with a formal collection of rewrite
rules~\cite{duncan2020graph}, so any diagram may be transformed into
structurally equivalent ones. This is what allows a synthesized circuit to be
simplified by axiomatic graph rewriting rather than by pattern-matching on a
gate list.

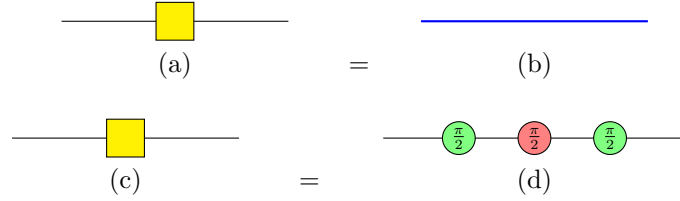
\begin{figure}[!ht]
\centering
  \begin{minipage}[b]{0.28\textwidth}
    \centering
    \begin{tikzpicture}[baseline=-0.9ex]
      \draw (-1.5,0) -- (1.5,0);
      \draw[fill=yellow, draw=black] (-0.25,-0.25) rectangle (0.25,0.25);
    \end{tikzpicture}
    \par\smallskip (a)
  \end{minipage}
  \quad=\quad
  \begin{minipage}[b]{0.28\textwidth}
    \centering
    \begin{tikzpicture}[baseline=-0.9ex]
      \draw[blue, thick] (-1.5,0) -- (1.5,0);
    \end{tikzpicture}
    \par\smallskip (b)
  \end{minipage}
  
  \vspace{0.5cm} 
  
  \begin{minipage}[b]{0.28\textwidth}
    \centering
    \begin{tikzpicture}[baseline=-0.9ex]
      \draw (-1.5,0) -- (1.5,0);
      \draw[fill=yellow, draw=black] (-0.25,-0.25) rectangle (0.25,0.25);
    \end{tikzpicture}
    \par\smallskip (c)
  \end{minipage}
  \quad=\quad
  \begin{minipage}[b]{0.38\textwidth}
    \centering
    \begin{tikzpicture}[baseline=-0.9ex]
      \draw (-1,0) -- (3,0);
      \node[fill=green!50, circle, draw, minimum size=4mm, inner sep=1pt] (A) at (0,0) {\scalebox{0.75}{$\frac{\pi}{2}$}};
      \node[fill=red!50, circle, draw, minimum size=4mm, inner sep=1pt] (B) at (1,0) {\scalebox{0.75}{$\frac{\pi}{2}$}};
      \node[fill=green!50, circle, draw, minimum size=4mm, inner sep=1pt] (C) at (2,0) {\scalebox{0.75}{$\frac{\pi}{2}$}};
    \end{tikzpicture}
    \par\smallskip (d)
  \end{minipage}
\caption{ZX-calculus representations: (\textbf{a},\textbf{b}) Alternative notation for the Hadamard gate (Hadamard edge); (\textbf{c},\textbf{d}) Euler decomposition rule.}
\label{fig:zx-combined}
\end{figure}

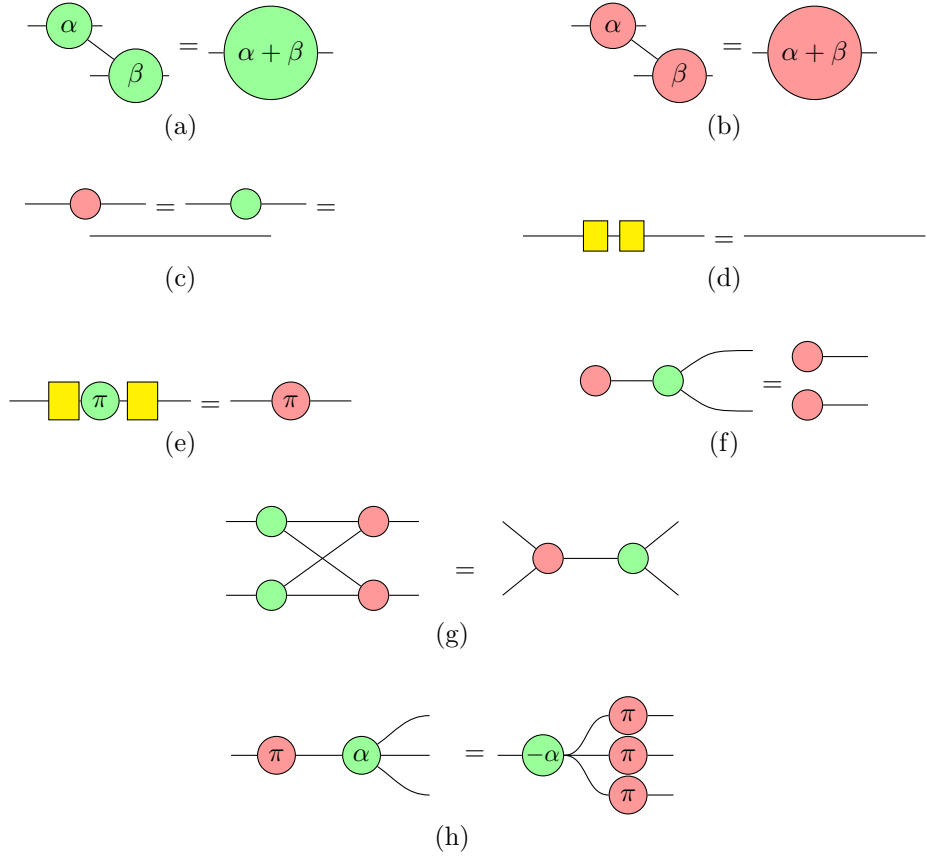
\begin{figure}[!ht]
\centering
  \begin{minipage}[b]{0.45\textwidth}
    \centering
    \begin{tikzpicture}[baseline=(current bounding box.center), scale=0.55]
        \node[circle, draw, fill=green!40, minimum size=6mm] (a) at (0,0.6) {$\alpha$};
        \node[circle, draw, fill=green!40, minimum size=6mm] (b) at (1.6,-0.6) {$\beta$};
        \draw (-1,0.6) -- (a);
        \draw (0.5,-0.6) -- (b);
        \draw (a) -- (b);
        \draw (a) -- ++(0.8,0);
        \draw (b) -- ++(0.8,0);
    \end{tikzpicture}
    $=$
    \begin{tikzpicture}[baseline=(current bounding box.center), scale=0.55]
        \node[draw, fill=green!40, circle, minimum size=6mm] (c) at (0,0) {$\alpha+\beta$};
        \draw (-1.5,0) -- (c) -- (1.5,0);
    \end{tikzpicture}
    \par\smallskip (a)
  \end{minipage}
  \hfill
  \begin{minipage}[b]{0.45\textwidth}
    \centering
    \begin{tikzpicture}[baseline=(current bounding box.center), scale=0.55]
        \node[circle, draw, fill=red!40, minimum size=6mm] (a) at (0,0.6) {$\alpha$};
        \node[circle, draw, fill=red!40, minimum size=6mm] (b) at (1.6,-0.6) {$\beta$};
        \draw (-1,0.6) -- (a);
        \draw (0.5,-0.6) -- (b);
        \draw (a) -- (b);
        \draw (a) -- ++(0.8,0);
        \draw (b) -- ++(0.8,0);
    \end{tikzpicture}
    $=$
    \begin{tikzpicture}[baseline=(current bounding box.center), scale=0.55]
        \node[draw, fill=red!40, circle, minimum size=6mm] (c) at (0,0) {$\alpha+\beta$};
        \draw (-1.5,0) -- (c) -- (1.5,0);
    \end{tikzpicture}
    \par\smallskip (b)
  \end{minipage}

  \vspace{0.6cm}

  \begin{minipage}[b]{0.45\textwidth}
    \centering
    \begin{tikzpicture}[baseline=-0.9ex, scale=0.8]
    \node[draw, fill=red!40, circle, minimum size=4mm] (a) at (0,0) {};
    \draw (-1,0) -- (a);
    \draw (a) -- (1,0);
    \end{tikzpicture}
    $=$
    \begin{tikzpicture}[baseline=-0.9ex, scale=0.8]
    \node[draw, fill=green!40, circle, minimum size=4mm] (c) at (0,0) {};
    \draw (-1,0) -- (c);
    \draw (c) -- (1,0);
    \end{tikzpicture}
    $=$
    \begin{tikzpicture}[baseline=-0.9ex, scale=0.8]
    \draw (-1,0) -- (2,0);
    \end{tikzpicture}
    \par\smallskip (c)
  \end{minipage}
  \hfill
  \begin{minipage}[b]{0.45\textwidth}
    \centering
    \begin{tikzpicture}[baseline=-0.9ex, scale=0.8]
    \draw (-1.5,0) -- (1.5,0);
    \draw[fill=yellow] (-0.5,-0.25) rectangle (-0.1,0.25);
    \draw[fill=yellow] (0.1,-0.25) rectangle (0.5,0.25);
    \end{tikzpicture}
    $=$ 
    \begin{tikzpicture}[baseline=-0.9ex, scale=0.8]
    \draw (-1,0) -- (2,0);
    \end{tikzpicture}
    \par\smallskip (d)
  \end{minipage}

  \vspace{0.6cm}

  \begin{minipage}[b]{0.45\textwidth}
    \centering
    \begin{tikzpicture}[baseline=-0.9ex, scale=0.8]
      \draw (-1.5,0) -- (-0.8,0);
      \draw (0.8,0) -- (1.5,0);
      \node[draw, fill=yellow, minimum width=4mm, minimum height=5mm] (a) at (-0.6,0) {};
      \node[draw, fill=green!40, circle, minimum size=5mm, inner sep=1pt] (b) at (0,0) {$\pi$};
      \node[draw, fill=yellow, minimum width=4mm, minimum height=5mm] (c) at (0.7,0) {};
      \draw (a) -- (b);
      \draw (b) -- (c);
    \end{tikzpicture}
    $=$ 
    \begin{tikzpicture}[baseline=-0.9ex, scale=0.8]
    \node[draw, fill=red!40, circle, minimum size=5mm, inner sep=1pt] (c) at (0,0) {$\pi$};
    \draw (-1,0) -- (c);
    \draw (c) -- (1,0);
    \end{tikzpicture}
    \par\smallskip (e)
  \end{minipage}
  \hfill
  \begin{minipage}[b]{0.45\textwidth}
    \centering
    \begin{tikzpicture}[baseline=-0.9ex, scale=0.8]
    \node[draw, fill=red!40, circle, minimum size=4mm] (r) at (0,0) {};
    \node[draw, fill=green!40, circle, minimum size=4mm] (g) at (1.2,0) {};
    \draw (r) -- (g);
    \draw (g) .. controls (1.9,0.5) .. (2.6,0.5);
    \draw (g) .. controls (1.9,-0.5) .. (2.6,-0.5);
    \end{tikzpicture}
    $=$
    \begin{tikzpicture}[baseline=-0.9ex, scale=0.8]
    \node[draw, fill=red!40, circle, minimum size=4mm] (r1) at (0,0.4) {};
    \node[draw, fill=red!40, circle, minimum size=4mm] (r2) at (0,-0.4) {};
    \draw (r1) -- (1,0.4);
    \draw (r2) -- (1,-0.4);
    \end{tikzpicture}
    \par\smallskip (f)
  \end{minipage}

  \vspace{0.6cm}

  \begin{minipage}[b]{1.0\textwidth}
    \centering
    \begin{tikzpicture}[baseline=-0.9ex, scale=0.75]
    \node[draw, fill=green!40, circle, minimum size=4mm] (g1) at (0,0.8) {};
    \node[draw, fill=green!40, circle, minimum size=4mm] (g2) at (0,-0.5) {};
    \node[draw, fill=red!40, circle, minimum size=4mm] (r1) at (1.8,0.8) {};
    \node[draw, fill=red!40, circle, minimum size=4mm] (r2) at (1.8,-0.5) {};
    \draw (-0.8,0.8) -- (g1);
    \draw (-0.8,-0.5) -- (g2);
    \draw (g1) -- (r1);
    \draw (r1) -- (2.6,0.8);
    \draw (r2) -- (2.6,-0.5);
    \draw (g2) -- (r2);
    \draw (g1) -- (r2);
    \draw (g2) -- (r1);
    \end{tikzpicture}
    \quad=\quad
    \begin{tikzpicture}[baseline=-0.9ex, scale=0.75]
    \node[draw, fill=red!40, circle, minimum size=4mm] (r) at (0,0.15) {};
    \node[draw, fill=green!40, circle, minimum size=4mm] (g) at (1.5,0.15) {};
    \draw (-0.8,0.8) -- (r);
    \draw (-0.8,-0.5) -- (r);
    \draw (g) -- (2.3,0.8);
    \draw (g) -- (2.3,-0.5);
    \draw (r) -- (g);
    \end{tikzpicture}
    \par\smallskip (g)
  \end{minipage}

  \vspace{0.6cm}

  \begin{minipage}[b]{1.0\textwidth}
    \centering
    \begin{tikzpicture}[baseline=-0.9ex, scale=0.75]
    \node[draw, fill=red!40, circle, minimum size=5mm, inner sep=1pt] (rpi) at (0,0) {$\pi$};
    \node[draw, fill=green!40, circle, minimum size=5mm, inner sep=1pt] (galpha) at (1.5,0) {$\alpha$};
    \draw (-0.8,0) -- (rpi);
    \draw (rpi) -- (galpha);
    \draw (galpha) to[out=35,in=180] (2.7,0.7);
    \draw (galpha) to[out=0,in=180]  (2.7,0);
    \draw (galpha) to[out=-35,in=180] (2.7,-0.7);
    \node at (3.5,0) {$=$};
    \node[draw, fill=green!40, circle, minimum size=5mm, inner sep=0.5pt] (gminus) at (4.7,0) {\scalebox{0.9}{$-\alpha$}};
    \node[draw, fill=red!40, circle, minimum size=5mm, inner sep=1pt] (r1) at (6.2,0.7) {$\pi$};
    \node[draw, fill=red!40, circle, minimum size=5mm, inner sep=1pt] (r2) at (6.2,0) {$\pi$};
    \node[draw, fill=red!40, circle, minimum size=5mm, inner sep=1pt] (r3) at (6.2,-0.7) {$\pi$};
    \draw (3.9,0) -- (gminus);
    \draw (gminus) to[out=0,in=180] (r1);
    \draw (gminus) to[out=0,in=180] (r2);
    \draw (gminus) to[out=0,in=180] (r3);
    \draw (r1) -- (7,0.7);
    \draw (r2) -- (7,0);
    \draw (r3) -- (7,-0.7);
    \end{tikzpicture}
    \par\smallskip (h)
  \end{minipage}
\caption{Fundamental rewrite rules of the ZX-calculus: (\textbf{a}) Green spider fusion; (\textbf{b}) red spider fusion; (\textbf{c}) identity rule 1; (\textbf{d}) identity rule 2; (\textbf{e}) color change rule; (\textbf{f}) copy rule; (\textbf{g}) bialgebra rule; (\textbf{h}) $\pi$ copy rule.}
\label{fig:zx-rules-all}
\end{figure}

Two of the rules in Figure~\ref{fig:zx-rules-all} do the work measured in this
paper. \emph{Spider fusion} (Figure~\ref{fig:zx-rules-all}a,b) merges adjacent
same-basis spiders and adds their phases; \emph{identity removal}
(Figure~\ref{fig:zx-rules-all}c,d) deletes a phase-free two-legged spider. Both
preserve the implemented linear map, so applying them cannot change the
approximation error of a synthesized circuit.

A diagram is \emph{graph-like} when all interior spiders are $Z$-spiders
joined exclusively by Hadamard edges, the underlying graph is simple, every
boundary vertex is attached to a $Z$-spider, and each $Z$-spider carries at
most one boundary~\cite{duncan2020graph}. Any circuit can be brought to this
form by the colour-change rule. Automated strategies additionally use local
complementation and pivoting to eliminate interior Clifford
spiders~\cite{duncan2020graph,duncan2013pivoting}; those operations are part
of the general theory but, as discussed in Section~\ref{sec:method-simp}, are
not what accounts for the reductions measured here.

\section{Numerical methodology}\label{sec:method}

The pipeline runs in four stages: synthesize, map to a ZX-diagram, simplify,
extract and measure.

\subsection{Benchmark set}\label{sec:method-target}

Targets were drawn at random from four families: the Pauli rotations
$R_x(\theta)$, $R_y(\theta)$ and $R_z(\theta)$ with $\theta$ uniform on
$[0, 2\pi)$, and the general single-qubit gate $U(\theta,\phi,\lambda)$ with all
three angles drawn independently from the same interval. One hundred targets
were generated per family at each of three recursion depths, giving $1200$
synthesis and optimization runs in total.

All targets are drawn in advance from a single seeded generator
(\texttt{numpy.random.default\_rng(20260809)}) and then dispatched, so that the
reported values do not depend on the order in which the runs are executed.

A single target, $R_x(2.3)$, is also carried through the whole pipeline as a
worked example, so that the behaviour at each depth can be followed in detail.
It is reported alongside the aggregates in Section~\ref{sec:results}, where it
serves to illustrate how far an individual target can sit from the mean.

\subsection{Synthesis}\label{sec:method-synth}

Each target was approximated with the \texttt{SolovayKitaev} transpiler pass of
Qiskit~2.5.1, over the inverse-closed gate set
$\{H, S, S^{\dagger}, T, T^{\dagger}, Z\}$ of Eq.~\eqref{eq:gateset}. The pass
implements the recursion of Section~\ref{sec:bg-sk} and takes the recursion
depth as its parameter; the base approximations it searches are those built into
the library, so no net was constructed for this work. Synthesis was repeated at
recursion depths $d = 2, 3, 4$.

Reporting the depth rather than a target $\epsilon$ is deliberate. The depth is
the parameter the algorithm actually takes, and the accuracy that follows from
it varies from target to target; the achieved error is therefore measured per
instance and reported as a mean rather than assumed from the convergence bound.
Averaged over the benchmark set, the trace error norm falls from
$8\times10^{-3}$ at $d = 2$ to $1.4\times10^{-3}$ at $d = 3$ and
$1\times10^{-4}$ at $d = 4$.

The distance between a target $U$ and a synthesized sequence $S$ is the trace
error norm~\cite{ozols2009solovay}
\begin{equation}\label{eq:tracenorm}
d(U, S) = \sqrt{\frac{2 - |\mathrm{Tr}(U^{\dagger}S)|}{2}} ,
\end{equation}
which vanishes when $S = U$ and, being a function of $|\mathrm{Tr}(U^\dagger S)|$
alone, is insensitive to a global phase. That last property matters here:
diagrammatic simplification discards the global scalar, so a phase-sensitive
metric such as the operator norm would report a spurious discrepancy between a
circuit and its optimized form even when the two implement the same operation.
Every error quoted in this paper is Eq.~\eqref{eq:tracenorm}.

\subsection{Mapping and simplification}\label{sec:method-simp}

The synthesized gate string is exported to QASM, parsed into a ZX-diagram, and
converted to graph-like form, with $S$ and $T$ becoming $Z$-spiders of phase
$\pi/2$ and $\pi/4$ and Hadamards becoming yellow
boxes~\cite{duncan2020graph,backens2014zx}. Simplification was performed with
PyZX~0.10.5~\cite{kissinger2019pyzx}, by calling \texttt{full\_reduce} on the
graph followed by \texttt{normalize}. \texttt{full\_reduce} applies the full
graph-theoretic strategy of Ref.~\cite{duncan2020graph}: spider fusion and
identity removal together with local complementation and pivoting to eliminate
interior Clifford spiders.

Because the rewrites preserve the implemented linear map, the approximation
error of the extracted circuit equals that of the synthesized circuit. This is a
property of the calculus rather than a finding, and Section~\ref{sec:res-tcount}
reports it as a check on the implementation.

\subsection{Extraction and metrics}\label{sec:method-metrics}

The simplified diagram is extracted back to a circuit with PyZX's
\texttt{extract\_circuit}~\cite{duncan2020graph}. Three quantities are recorded
before and after optimization: total gate count, $T$-count, and wall-clock time.

$T$-count is the number of gates carrying a phase that is an odd multiple of
$\pi/4$. Phases of $\pi/4$ and $7\pi/4$ are $T$ and $T^{\dagger}$; phases of
$3\pi/4$ and $5\pi/4$ factor as $S\!\cdot\!T$ and $Z\!\cdot\!T$ and so cost one
$T$ apiece. Clifford phases contribute nothing. Counting this way makes the
before and after figures comparable, since extraction is free to leave a
non-Clifford phase in any of these four forms.

Timings were measured in a separate serial run, on the same machine and with the
same code path but with one target processed at a time, since wall-clock
measurements taken while several targets are processed concurrently are
distorted by contention. Each timing brackets the synthesis call alone or the
map--simplify--extract sequence alone. The timing run extends one level below
the benchmark set, covering $d = 1, 2, 3, 4$ where the gate-count and
$T$-count measurements cover $d = 2, 3, 4$ only: $d = 1$ is too coarse to be
of interest as an approximation, but it locates the setting at which the
rewriting layer is still negligible against synthesis, which is what
Section~\ref{sec:res-runtime} is measuring. The pass caches its base approximations on
first use, so a warm-up call precedes the measured runs and the reported
synthesis times exclude that one-off construction; Section~\ref{sec:res-runtime}
returns to this point.

All runs were performed on the CPU runtime of Google Colab, an Intel Xeon at
$2.20$~GHz with $12$~GB of RAM, with no GPU acceleration. Nothing in the
pipeline is amenable to it: the work is a recursion over $2\times2$ matrices
followed by graph rewriting.

\section{Results}\label{sec:results}

\subsection{Total gate count}\label{sec:res-total}

Table~\ref{tab:bulk-total} reports the total gate footprint before and after
optimization, averaged over one hundred random targets in each of the four
families at each depth.

\begin{table}[!t]
\centering
\caption{Total gate count before and after ZX-calculus post-processing, over
$100$ random targets per family per depth. Counts are means; reductions are
means $\pm$ one standard deviation across targets.}
\label{tab:bulk-total}
\begin{tabular}{@{}llrrr@{}}
\toprule
Family & $d$ & Initial & Optimized & Reduction (\%) \\
\midrule
$R_x$ & 2 & 218.7 & 158.9 & $26.63 \pm 7.78$ \\
$R_y$ & 2 & 218.5 & 156.8 & $28.09 \pm 8.25$ \\
$R_z$ & 2 & 214.7 & 152.6 & $28.50 \pm 7.08$ \\
$U(\theta,\phi,\lambda)$ & 2 & 233.4 & 168.0 & $27.94 \pm 7.13$ \\
\midrule
$R_x$ & 3 & 1130.1 & 806.8 & $28.66 \pm 5.08$ \\
$R_y$ & 3 & 1140.1 & 810.7 & $28.91 \pm 5.66$ \\
$R_z$ & 3 & 1122.5 & 782.8 & $30.13 \pm 5.57$ \\
$U(\theta,\phi,\lambda)$ & 3 & 1141.5 & 818.3 & $28.29 \pm 5.04$ \\
\midrule
$R_x$ & 4 & 5567.2 & 3953.1 & $28.97 \pm 2.70$ \\
$R_y$ & 4 & 5612.8 & 3986.5 & $28.99 \pm 3.39$ \\
$R_z$ & 4 & 5621.0 & 3993.4 & $28.96 \pm 3.29$ \\
$U(\theta,\phi,\lambda)$ & 4 & 5544.3 & 3951.9 & $28.73 \pm 3.28$ \\
\bottomrule
\end{tabular}
\end{table}

Deeper recursion lengthens the synthesized word steeply: the mean grows from
about $220$ gates at $d = 2$ to about $1130$ at $d = 3$ and about $5580$ at
$d = 4$, a factor of roughly five per level, as Eq.~\eqref{eq:length} leads one
to expect. The recursion appends a further layer of nested commutators at each
level in order to meet a stricter proximity requirement, without regard to the
length or local economy of the resulting instruction string.

Post-processing removes a little under a third of that in every configuration
tested. Three features of the table are worth separating.

The fractional reduction is close to constant in the recursion depth. It rises
from $d = 2$ to $d = 3$ in all four families, by between $0.35$ and $2.03$
percentage points, and is then flat from $d = 3$ to $d = 4$. Over a range in
which the circuit lengthens twenty-five-fold, the proportion of it that
rewriting removes moves by at most about two percentage points.

The fractional reduction is also close to constant across the four families,
and increasingly so with depth. At $d = 2$ the family means span $1.9$
percentage points; at $d = 4$ they span $0.26$ percentage points, against a
standard error of the mean of about $0.3$, so at that depth the four families
are not distinguishable from one another. Whatever rewriting is recovering, it
does not depend on which rotation axis the target lies on.

The spread across targets collapses as the depth increases: the standard
deviation of the reduction falls from $7.1$--$8.3$ percentage points at
$d = 2$ to $5.0$--$5.7$ at $d = 3$ and $2.7$--$3.4$ at $d = 4$. Shallow
synthesis is idiosyncratic, and individual targets can depart substantially from
the mean; deep synthesis is not, and the reduction becomes a stable property of
the construction rather than of the particular target.

That last point bears directly on how a single target should be read. The worked
example $R_x(2.3)$ yields reductions of $31.11\%$, $28.19\%$ and $28.94\%$ at
$d = 2, 3, 4$, from initial counts of $180$, $1064$ and $5812$ gates. At $d = 3$
and $d = 4$ those figures sit essentially on the corresponding means. At
$d = 2$, however, the target is a favourable draw: its reduction lies about
$0.6$ standard deviations above the $R_x$ mean and its synthesized word is
shorter than average. Read on its own, the sequence $31.11 \rightarrow 28.19
\rightarrow 28.94$ suggests that the margin narrows with depth. The aggregate
shows that it does not; the apparent narrowing is an artefact of one favourable
point at the shallowest setting. We report this because it is the kind of
inference a single-target study invites, and because it is the reason the
benchmark was run.

\begin{figure}[!ht]
\centering
\includegraphics[width=0.95\textwidth]{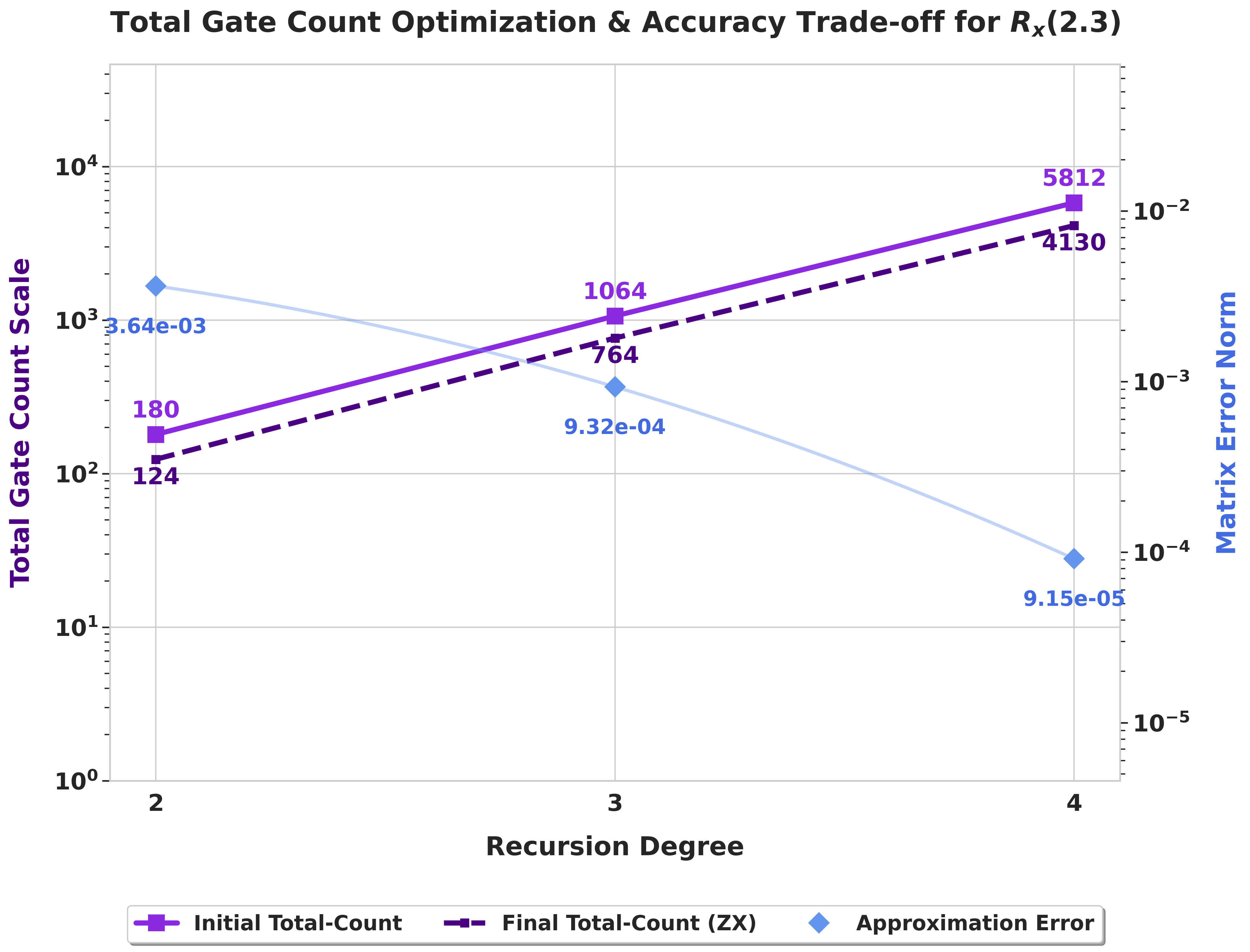}
\caption{Total gate count before and after ZX-calculus post-processing for the
worked example $R_x(2.3)$, with the achieved error on the right-hand axis,
against recursion depth. The error series is a function of the depth alone and
is unchanged by the optimization.}
\label{fig:total}
\end{figure}

\begin{figure}[!ht]
\centering
\includegraphics[width=0.95\textwidth]{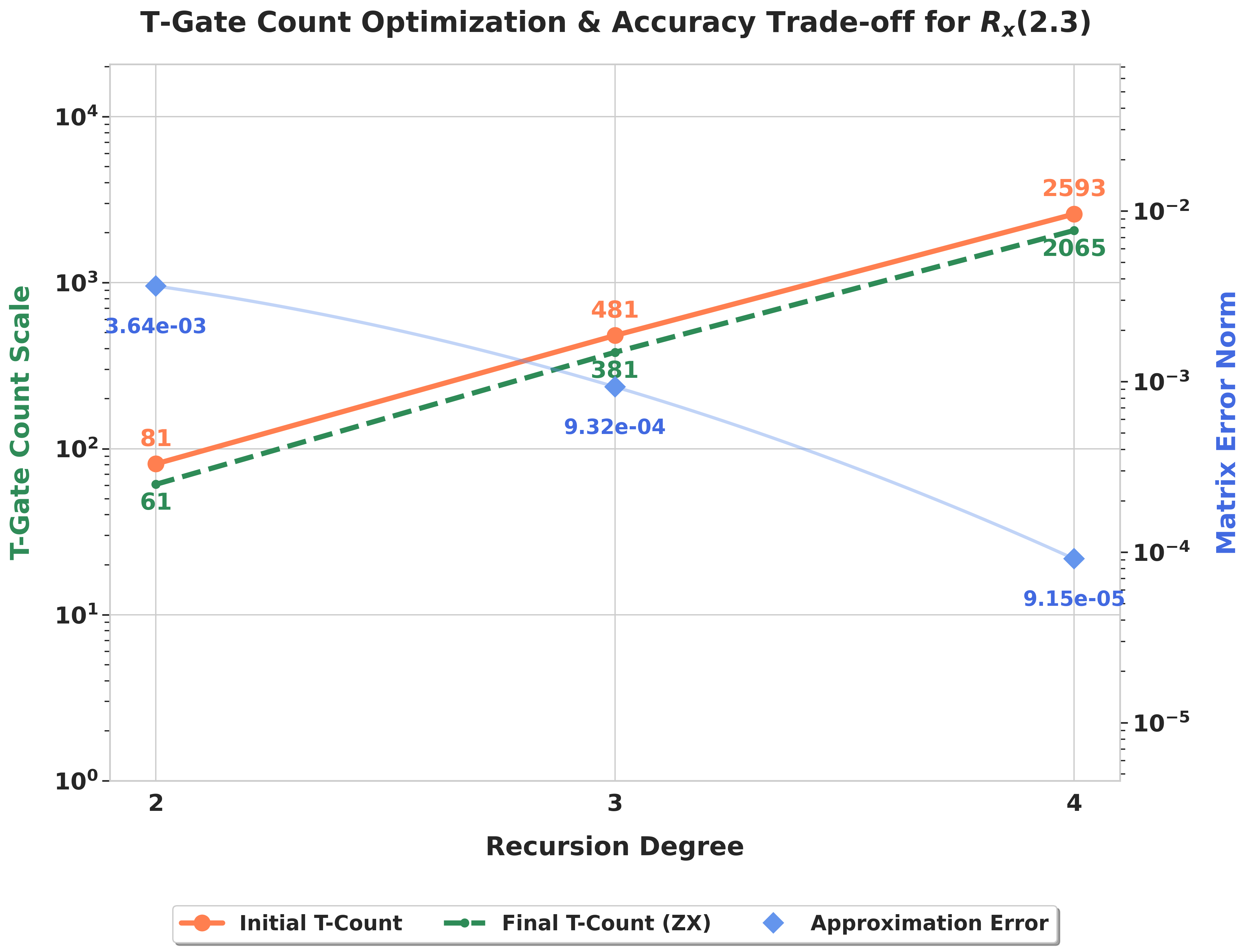}
\caption{$T$-count before and after ZX-calculus post-processing for the worked
example $R_x(2.3)$, with the achieved error on the right-hand axis, against
recursion depth.}
\label{fig:tcount}
\end{figure}

The worked example is shown in Figures~\ref{fig:total} and~\ref{fig:tcount},
which plot its gate counts and achieved error against the recursion depth.

While the fraction stays flat, the absolute saving grows with the circuit: on
average $60$--$65$ gates are removed at $d = 2$, $323$--$340$ at $d = 3$, and
$1592$--$1628$ at $d = 4$.

The mechanism is what the two rules of Section~\ref{sec:bg-zx} would suggest.
Reading the circuit as a diagram discards the temporal ordering, so co-axial
phase spiders separated in the gate string by intervening basis changes become
adjacent in the graph and fuse, and the phase-free two-legged spiders that
result are removed as identities. A linear compiler scanning the gate list
cannot bring those phases together. That this recovers a roughly fixed share of
the output, rather than a growing one, is consistent with the redundancy being
generated locally at the junctions between the base words that the recursion
concatenates, rather than accumulating with the nesting depth.

\subsection{\texorpdfstring{$T$}{T}-count}\label{sec:res-tcount}

Table~\ref{tab:bulk-t} reports the non-Clifford footprint, which is the quantity
that governs distillation cost.

\begin{table}[!t]
\centering
\caption{$T$-count before and after ZX-calculus post-processing, over $100$
random targets per family per depth. Counts are means; reductions are means
$\pm$ one standard deviation across targets.}
\label{tab:bulk-t}
\begin{tabular}{@{}llrrr@{}}
\toprule
Family & $d$ & Initial & Optimized & Reduction (\%) \\
\midrule
$R_x$ & 2 & 97.6 & 78.9 & $18.51 \pm 7.82$ \\
$R_y$ & 2 & 96.8 & 77.8 & $19.59 \pm 9.13$ \\
$R_z$ & 2 & 95.4 & 75.9 & $20.18 \pm 7.62$ \\
$U(\theta,\phi,\lambda)$ & 2 & 104.8 & 83.6 & $20.21 \pm 8.01$ \\
\midrule
$R_x$ & 3 & 508.7 & 403.0 & $20.87 \pm 5.63$ \\
$R_y$ & 3 & 513.4 & 404.6 & $21.21 \pm 6.38$ \\
$R_z$ & 3 & 503.1 & 390.9 & $22.22 \pm 5.81$ \\
$U(\theta,\phi,\lambda)$ & 3 & 514.9 & 408.8 & $20.63 \pm 5.31$ \\
\midrule
$R_x$ & 4 & 2509.6 & 1976.0 & $21.26 \pm 2.83$ \\
$R_y$ & 4 & 2526.9 & 1992.6 & $21.18 \pm 3.62$ \\
$R_z$ & 4 & 2529.3 & 1996.3 & $21.09 \pm 3.46$ \\
$U(\theta,\phi,\lambda)$ & 4 & 2494.6 & 1975.6 & $20.83 \pm 3.50$ \\
\bottomrule
\end{tabular}
\end{table}

The unoptimized $T$-count tracks the total gate count at a stable share of
roughly $45\%$ throughout. Post-processing reduces it by about a fifth, and the
pattern of Section~\ref{sec:res-total} recurs: the fraction rises from $d = 2$
to $d = 3$ in all four families, is flat thereafter, and converges to
$20.8$--$21.3\%$ across families at $d = 4$, with the standard deviation falling
from $7.6$--$9.1$ percentage points to $2.8$--$3.6$. In absolute terms the
saving grows from about $20$ $T$ gates to about $530$.

The $T$-count reduction is consistently smaller than the total gate reduction,
by between $7.7$ and $8.5$ percentage points. This is expected: fusing two adjacent
$\pi/4$ spiders yields a single Clifford $\pi/2$ spider, which removes two $T$
gates but only one gate overall, so the two metrics respond differently to the
same rewrite.

Because the rewrites of Section~\ref{sec:bg-zx} preserve the implemented linear
map, this reduction costs nothing in accuracy. That was verified rather than
assumed. For every one of the $1200$ runs the operator of the extracted circuit
was compared with that of the synthesized circuit through
$1 - |\mathrm{Tr}(U_{\mathrm{synth}}^{\dagger} U_{\mathrm{opt}})|/2$, a quantity
that vanishes when the two agree up to a global phase. The largest value
observed was $5.8 \times 10^{-13}$, which is floating-point accumulation over
several thousand gate products. The optimized circuit implements the same
operator as the synthesized one, and the errors quoted in
Section~\ref{sec:method-synth} apply unchanged after optimization.

\subsection{Compile-time cost}\label{sec:res-runtime}

\begin{figure}[!ht]
\centering
\includegraphics[width=0.85\textwidth]{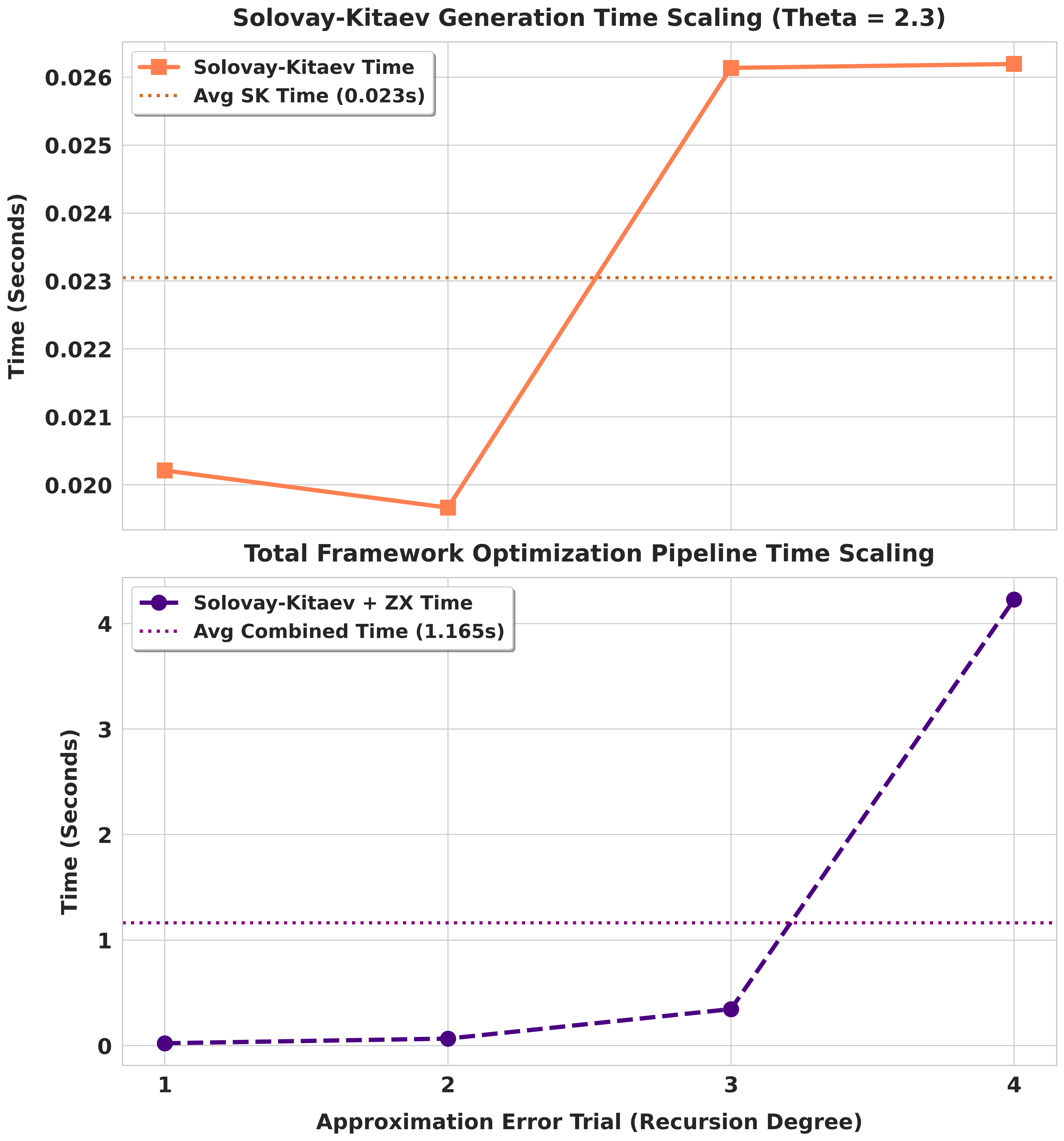}
\caption{Wall-clock time against recursion depth for the worked example. Upper
panel: synthesis alone. Lower panel: the combined pipeline. The dotted lines
mark the means over the four depths.}
\label{fig:runtime}
\end{figure}

A structural optimizer earns its place only if the cost of running it is
tolerable. Table~\ref{tab:runtime} separates the time spent in synthesis from
the time spent in rewriting for the worked example, and Figure~\ref{fig:runtime}
plots both.

\begin{table}[!t]
\centering
\caption{Wall-clock time for Solovay--Kitaev synthesis alone and for the
combined pipeline, for the worked example $R_x(2.3)$, with the marginal cost of
the rewriting layer.}
\label{tab:runtime}
\begin{tabular}{@{}lrrr@{}}
\toprule
Depth & Synthesis (s) & Combined (s) & Marginal ZX cost (s) \\
\midrule
$d = 1$ & 0.0202 & 0.0235 & 0.0033 \\
$d = 2$ & 0.0197 & 0.0658 & 0.0461 \\
$d = 3$ & 0.0261 & 0.3440 & 0.3179 \\
$d = 4$ & 0.0262 & 4.2280 & 4.2018 \\
\midrule
Mean    & 0.0231 & 1.1653 & 1.1423 \\
\bottomrule
\end{tabular}
\end{table}

The two costs behave quite differently. The synthesis column is nearly flat,
moving only from $0.0202$~s to $0.0262$~s across the four depths, while the
marginal cost of the rewriting layer grows from $0.0033$~s to $4.2018$~s, a
factor of about $1270$.

The flatness of the synthesis column requires a word of explanation, because it
is not the cost of the recursion. Each figure in that column includes the
one-off construction of the library's base approximations, which is performed on
first use and cached, and which does not depend on the recursion depth; at these
depths it dominates the measurement. Timed separately, with the cache already
warm, synthesis takes $0.8$~ms at $d = 2$ and $1.9$~ms at $d = 3$ over the
benchmark set, so the recursion itself does grow with depth, by rather more than
a factor of two per level. What Table~\ref{tab:runtime} shows is that both
components of synthesis remain small in absolute terms next to the rewriting
layer, not that the recursion is free.

The rewriting cost is genuine and grows steeply. Measured over the benchmark set
it rises from $41$~ms at $d = 2$ to $335$~ms at $d = 3$, an eightfold increase
for a fivefold increase in circuit length, and the worked example puts $d = 4$
at $4.2$~s. Since the circuit lengthens by about a factor of five per level, the
rewriting cost is growing superlinearly in circuit size.

The practical reading is that the composition has an operating range rather than
a uniform benefit. At shallow depth the rewriting layer is essentially free
relative to synthesis. At $d = 4$ it accounts for over $99\%$ of the total
pipeline time and dominates it outright. Against that, the cost is paid once at
compile time whereas the gate saving is realized on every execution of the
circuit, so the trade is favourable for any circuit that will be run more than a
handful of times; but it is a trade, and it should not be described as a free
improvement.

\subsection{Limitations}\label{sec:limitations}

Three limitations bound what the measurements above support, and are stated here
rather than left for a reader to infer.

The comparison is against unoptimized Solovay--Kitaev output only. No comparison
is made against other optimizers, so the results do not establish that the
reduction is specific to diagrammatic rewriting rather than available to any
peephole optimizer, and no comparison is made against near-optimal synthesis,
against which the absolute $T$-counts here are not competitive (see
Section~\ref{sec:related-synth}).

The reductions are a property of the particular simplification strategy invoked,
\texttt{full\_reduce} as shipped in PyZX~0.10.5. A different strategy, or a
later version, would give different numbers.

The timings in Section~\ref{sec:res-runtime} are single measurements on one
target and should be read as indicative of magnitude rather than as precise
figures; the benchmark-set means quoted alongside them are averages over one
hundred targets and are firmer.

\section{Conclusion}\label{sec:conclusion}

The $T$-count of a synthesized circuit dominates its cost under fault
tolerance, because each $T$ gate is paid for in distillation. The
Solovay--Kitaev construction meets an accuracy target with a sequence length
that is polylogarithmic in the inverse error, but it optimizes for convergence
and leaves structural redundancy behind. This paper measured how much of that
redundancy automated ZX-calculus rewriting recovers, over twelve hundred random
single-qubit targets at three recursion depths.

Post-processing removed $26.6$--$30.1\%$ of the total gate count and
$18.5$--$22.2\%$ of the $T$-count, at no cost in accuracy: the rewrite rules
preserve the implemented linear map, and the extracted circuit was verified
against the synthesized one on every run, agreeing to within
$5.8 \times 10^{-13}$.

Three features of the result seem worth carrying forward. The fraction removed
is close to constant in the recursion depth, rising slightly from the shallowest
setting and then flat, while the absolute saving grows from about $60$ gates to
about $1600$; the recoverable redundancy therefore appears to be a roughly fixed
share of the output rather than something that accumulates with nesting. That
fraction is also close to constant across target families, and the four families
become statistically indistinguishable at the deepest setting while the spread
across individual targets falls sharply, which suggests the quantity being
measured is a property of the construction rather than of the target. And the
compile-time cost of recovering it grows steeply, from negligible at $d = 1$ to
dominant at $d = 4$, so the composition should be presented as a trade with an
operating range and not as a free improvement.

\subsection*{Future work}

Three directions follow. The first is breadth of comparison, both against
alternative inverse-closed generating sets and against non-diagrammatic
optimizers, which would separate what is specific to ZX rewriting from what any
optimizer recovers. The second is depth: the flat behaviour reported here covers
three recursion levels, and whether the fraction remains constant as the
accuracy target is tightened further is a question the present data cannot
settle, though the collapse of the spread with depth suggests it will. The third
is the multi-qubit setting, where the analogous question concerns entangling
structure and CNOT count rather than single-qubit phases, and where the
compile-time behaviour reported in Section~\ref{sec:res-runtime} is likely to be
the binding constraint.

\backmatter

\section*{Declarations}

\bmhead{Funding}
C.-F.K. is grateful to the European Union and the Region Reunion, France
(POE FEDER 2021--2027, n$^\circ$2025-0954-007180) for the funding support.
K.D.S. is grateful to the European Union and the Region Reunion, France
(POE FEDER 2021--2027, n$^\circ$2025-0963-007179) for the funding support.
A.M. is grateful to the European Union and the Region Reunion, France
(POE FEDER 2021--2027, n$^\circ$2026-0237-010722) for the funding support.
PEACCEL is supported through a research program partially cofunded by the
European Union (UE) and Region Reunion (FEDER).

\bmhead{Competing interests}
The authors have no competing interests to declare that are relevant to the
content of this article.

\bmhead{Data availability}
The measured data supporting the results reported in this article, together with
the notebooks and the benchmark script that produced them, are openly available
in Zenodo at \texttt{https://doi.org/10.5281/zenodo.\DOI}. This comprises the
per-run records and the per-configuration summaries behind
Tables~\ref{tab:bulk-total} and~\ref{tab:bulk-t}, and the notebooks behind
Table~\ref{tab:runtime} and Figures~\ref{fig:total}--\ref{fig:runtime}.

\bmhead{Code availability}
The synthesis, optimization and benchmarking code is openly available in the
same Zenodo record, \texttt{https://doi.org/10.5281/zenodo.\DOI}, which documents
the package versions and the random seed required to reproduce the reported
values.

\bmhead{Author contributions}
A. Mahasinghe conceived the project, developed the theoretical framework and
supervised the work. D. De Silva implemented the synthesis and optimization
pipeline, performed the numerical experiments for the worked example reported in
Section~\ref{sec:res-runtime} and Figures~\ref{fig:total}--\ref{fig:runtime},
prepared Figures~\ref{fig:zx_spiders}--\ref{fig:zx-rules-all}, and wrote the
first draft of the manuscript. C.-F. Kam carried out the benchmark reported in
Tables~\ref{tab:bulk-total} and~\ref{tab:bulk-t} and the verification described
in Section~\ref{sec:res-tcount}, revised the theoretical presentation in
Section~\ref{sec:background}, and restructured and redrafted the manuscript for
submission. K. De Silva advised on the study during discussions and reviewed the
manuscript. F. Cadet and J. Wang advised on the work, and reviewed and edited the
manuscript. All authors read and approved the submitted version.

\bmhead{Ethics approval}
Not applicable; this study involves neither human participants nor animals.

\bmhead{Use of AI tools}
A large language model was used to restructure and redraft the manuscript text
following an internal pre-submission review. All technical content, all
numerical results and all figures are the authors' own; no result, table or
figure was generated by the model. The authors take full responsibility for the
final text.


\begin{thebibliography}{99}

\bibitem{nielsen2010quantum}
M.A. Nielsen, I.L. Chuang, \emph{Quantum Computation and Quantum Information},
10th anniversary edn. (Cambridge University Press, Cambridge, 2010)

\bibitem{kitaev1997quantum}
A.Y. Kitaev, Quantum computations: algorithms and error correction. Russ.
Math. Surv. \textbf{52}(6), 1191--1249 (1997).
\url{https://doi.org/10.1070/RM1997v052n06ABEH002155}

\bibitem{kitaev2002classical}
A.Y. Kitaev, A. Shen, M.N. Vyalyi, \emph{Classical and Quantum Computation}.
Graduate Studies in Mathematics, vol. 47 (American Mathematical Society,
Providence, 2002)

\bibitem{nielsen2001quantum}
M.A. Nielsen, I.L. Chuang, \emph{Quantum Computation and Quantum Information},
1st edn. (Cambridge University Press, Cambridge, 2000)

\bibitem{aaronson2004improved}
S. Aaronson, D. Gottesman, Improved simulation of stabilizer circuits. Phys.
Rev. A \textbf{70}, 052328 (2004).
\url{https://doi.org/10.1103/PhysRevA.70.052328}

\bibitem{amy2014polynomial}
M. Amy, D. Maslov, M. Mosca, Polynomial-time $T$-depth optimization of
Clifford$+T$ circuits via matroid partitioning. IEEE Trans. Comput.-Aided Des.
Integr. Circuits Syst. \textbf{33}(10), 1476--1489 (2014).
\url{https://doi.org/10.1109/TCAD.2014.2341953}

\bibitem{kissinger2019reducing}
A. Kissinger, J. van de Wetering, Reducing the number of non-Clifford gates in
quantum circuits. Phys. Rev. A \textbf{102}, 022406 (2020).
\url{https://doi.org/10.1103/PhysRevA.102.022406}

\bibitem{dawson2005solovay}
C.M. Dawson, M.A. Nielsen, The Solovay--Kitaev algorithm. Quantum Inf. Comput.
\textbf{6}(1), 81--95 (2006). arXiv:quant-ph/0505030

\bibitem{mahasinghe2020solovay}
A. Mahasinghe, S. Bandaranayake, K. De Silva, Solovay--Kitaev approximations of
special orthogonal matrices. Adv. Math. Phys. \textbf{2020}, 2530609 (2020).
\url{https://doi.org/10.1155/2020/2530609}

\bibitem{ozols2009solovay}
M. Ozols, The Solovay--Kitaev theorem. Essay, University of Waterloo (2009)

\bibitem{roje2019solovay}
D. Roje, The Solovay--Kitaev theorem (2019)

\bibitem{kmm2013}
V. Kliuchnikov, D. Maslov, M. Mosca, Asymptotically optimal approximation of
single qubit unitaries by Clifford and $T$ circuits using a constant number of
ancillary qubits. Phys. Rev. Lett. \textbf{110}, 190502 (2013).
\url{https://doi.org/10.1103/PhysRevLett.110.190502}

\bibitem{ross-selinger2016}
N.J. Ross, P. Selinger, Optimal ancilla-free Clifford$+T$ approximation of
$z$-rotations. Quantum Inf. Comput. \textbf{16}(11--12), 901--953 (2016).
arXiv:1403.2975

\bibitem{coecke2011interacting}
B. Coecke, R. Duncan, Interacting quantum observables: categorical algebra and
diagrammatics. New J. Phys. \textbf{13}, 043016 (2011).
\url{https://doi.org/10.1088/1367-2630/13/4/043016}

\bibitem{coecke2018picturing}
B. Coecke, A. Kissinger, \emph{Picturing Quantum Processes} (Cambridge
University Press, Cambridge, 2017)

\bibitem{duncan2020graph}
R. Duncan, A. Kissinger, S. Perdrix, J. van de Wetering, Graph-theoretic
simplification of quantum circuits with the ZX-calculus. Quantum \textbf{4},
279 (2020). \url{https://doi.org/10.22331/q-2020-06-04-279}

\bibitem{backens2014zx}
M. Backens, The ZX-calculus is complete for stabilizer quantum mechanics. New
J. Phys. \textbf{16}, 093021 (2014).
\url{https://doi.org/10.1088/1367-2630/16/9/093021}

\bibitem{duncan2013pivoting}
R. Duncan, S. Perdrix, Pivoting makes the ZX-calculus complete for real
stabilizers. Electron. Proc. Theor. Comput. Sci. \textbf{171}, 50--62 (2014).
\url{https://doi.org/10.4204/EPTCS.171.5}

\bibitem{kissinger2019pyzx}
A. Kissinger, J. van de Wetering, PyZX: large scale automated diagrammatic
reasoning. Electron. Proc. Theor. Comput. Sci. \textbf{318}, 229--241 (2020).
\url{https://doi.org/10.4204/EPTCS.318.14}

\bibitem{kissinger2019cnot}
A. Kissinger, A. Meijer-van de Griend, CNOT circuit extraction for
topologically-constrained quantum memories. Quantum Inf. Comput.
\textbf{20}(7--8), 581--596 (2020). arXiv:1904.00633

\bibitem{nam2018automated}
Y. Nam, N.J. Ross, Y. Su, A.M. Childs, D. Maslov, Automated optimization of
large quantum circuits with continuous parameters. npj Quantum Inf.
\textbf{4}, 23 (2018). \url{https://doi.org/10.1038/s41534-018-0072-4}

\bibitem{zhang2019optimizing}
F. Zhang, J. Chen, Optimizing $T$ gates in Clifford$+T$ circuit as $\pi/4$
rotations around Paulis (2019). arXiv:1903.12456

\end{thebibliography}
\end{document}